\documentclass[aps,prl,twocolumn,showpacs,superscriptaddress]{revtex4}

\usepackage{graphicx}
\usepackage{amsmath}

\begin{document}

\title{Instability and Evolution of Nonlinearly Interacting Water Waves}

\author{P. K. Shukla}
\affiliation{Centre for Nonlinear
Physics, Department of Physics, Ume\aa~University, SE-90187
Ume\aa, Sweden}
\affiliation{Institut f\"ur Theoretische
Physik IV and Centre for Plasma Science and Astrophysics,
Fakult\"at f\"ur Physik und Astronomie, Ruhr--Universit\"at
Bochum, D-44780 Bochum, Germany}

\author{I. Kourakis}
\affiliation{Institut f\"ur Theoretische
Physik IV and Centre for Plasma Science and Astrophysics,
Fakult\"at f\"ur Physik und Astronomie, Ruhr--Universit\"at
Bochum, D-44780 Bochum, Germany}

\author{B. Eliasson}
\affiliation{Institut f\"ur Theoretische
Physik IV and Centre for Plasma Science and Astrophysics,
Fakult\"at f\"ur Physik und Astronomie, Ruhr--Universit\"at
Bochum, D-44780 Bochum, Germany}

\author{M. Marklund}
\affiliation{Centre for Nonlinear
Physics, Department of Physics, Ume\aa~University, SE-90187
Ume\aa, Sweden}

\author{L. Stenflo}
\affiliation{Centre for Nonlinear
Physics, Department of Physics, Ume\aa~University, SE-90187
Ume\aa, Sweden}

\date{Received 16 February 2006, revised 3 August 2006}

\begin{abstract}
We consider the modulational instability of nonlinearly interacting two-dimensional
waves in deep water, which are described by a pair of two-dimensional coupled
nonlinear Schr\"odinger equations. We derive a nonlinear dispersion relation. 
The latter is numerically analyzed to obtain the regions and the associated growth 
rates of the modulational instability. Furthermore, we follow the long term 
evolution of the latter by means of computer simulations of the governing 
nonlinear equations and demonstrate the formation of localized coherent wave envelopes.
Our results should be useful for understanding the formation and nonlinear propagation 
characteristics of large amplitude freak waves in deep water.
\end{abstract}

\pacs{47.35.-i,92.10.Hm}

\maketitle

Recently, there has been much interest \cite{r1,r2,r3,r4,r5} in investigating
the nonlinear formation of freak waves (also known as rogue waves, killer waves,
or giant waves) in the oceans. Such waves are responsible for the
loss of many ships and lives \cite{Kharif03}. Ocean waves usually have a spectrum of waves
that depends on the variations in the wind speed and direction \cite{Hasselmann76,Hasselmann80}. 
Freak waves are extraordinarily large amplitude localized water surface excitations, whose heights 
exceed many times the wavetrain height \cite{r6}. They may occur both in deep and shallow water, 
and may be created due to both statistical, linear (geometrical and spatio-temporal focusing) 
and nonlinear effects \cite{Kharif03}.  In nonlinear models, these steep objects are created 
on account of a delicate balance between the nonlinearity of the fluid and wave dispersion. The 
nonlinear dynamics of a single modulated large-amplitude water wavetrain can be modeled by a nonlinear
Schr\"odinger equation \cite{r7,r8,r8a,r8b,r8c}, which admits the Benjamin-Feir (modulational)
instability \cite{r9,r10,r11,Segur05} and the formation of rogue waves \cite{Calini02}. 
Even though a wide spectrum in the wave field may stabilize a wave train,
the Benjamin-Feir instability is quite robust against a narrow 
spectrum random field \cite{Kharif03}; it occurs if the spectral width $\sigma$ of the wave
is less than twice the average steepness (defined as $k_0 \bar{a}$ where $k_0$
is the central wavenumber and $\bar{a}$ is the rms value of the amplitude) \cite{Dysthe03}.
We stress that collapse of waves is inherently a two-dimensional problem, and that the single 
wave case has been investigated extensively in Ref.\ \cite{r11}. This indicates that the higher 
dimensionality of the problem at hand introduces a new flavor to the dynamics of interacting deep 
water waves, and brings us closer to observational prerequisites for such waves \cite{r11a}.

Onorato {\it et al.} \cite{r3} recently developed a simple weakly nonlinear
model for two nonlinearly interacting water waves in deep water
with two different directions of propagation. They showed that the
dynamics of these coupled waves is governed by two coupled nonlinear Schr\"odinger (CNLS)
equations, and presented an investigation of the modulational
instability for a one-dimensional two-wave system.
A fourth-order nonlinear evolution equation for two Stoke wave trains in deep water
was derived by Dhar and Das \cite{Dhar91}.  The two-wave system is similar
to what has also been investigated in nonlinear optics \cite{r12,r12a}, in Bose-Einstein
condensates \cite{r13, r13a}, in signal transmission lines \cite{r14}, in nonlinear
negative refraction index meta-materials \cite{r15,marklund-etal}, and in nonlinear
plasmas \cite{Inoue,r16, r16a, r16b,vladimirov-etal}.

In this Letter, we shall present for the first time a theory for the
modulational instability of a pair of two-dimensional 
nonlinearly coupled water waves in deep water, as well as the formation and dynamics 
of localized freak wave packets. For this purpose,
we shall use the CNLS equations of Onorato {\it et al.}\ \cite{r3},
which are valid for a system of obliquely propagating waves
(crossing sea states). In Ref.\ \cite{r3} the $x-$axis has been
defined as the middle (dichotome) between the two directions of
propagation, viz. $\mathbf{k}_A = (k_{A, x}, k_{A, y}) \equiv (k,
l)$ and $\mathbf{k}_B = (k_{B, x}, k_{B, y}) \equiv (k, -l)$; we
shall assume that both $k$ and $l$ are larger than zero.
The frequencies $\omega_j$ of the two carrier waves (i.e. $j = A,\, B$) are
related to the wavevectors $\mathbf{k}_j$ by the deep water dispersion relation \cite{Karpman}
$\omega_j = \sqrt{g|\mathbf{k}_j|}$, where $g$ denotes 
the gravitational acceleration. We note that the above hypothesis on the two wave
propagation directions implies $\omega_A = \omega_B = \sqrt{g\kappa}$,
where we have defined the wavenumber norm $\kappa \equiv \sqrt{k^2 + l^2}$. 
Moreover, we investigate the full dynamics of nonlinearly interacting deep water 
waves subjected to modulational/filamentation instabilities. It is found that 
random perturbations can grow to form inherently nonlinear water wave structures, 
the so called freak waves, through the nonlinear interaction between
two coupled water waves.  The latter should be of interest for explaining recent 
observations in water wave dynamics.  

Multiplying by $i$ the system of two-dimensional CNLS equations (4) and (5) of Ref.\ \cite{r3}, and
correcting a misprint (the coefficient in front of the $\partial^2B/\partial x\partial y$
term in Eq. (5) of Ref.\ \onlinecite{r3}), we have
\begin{subequations}
\begin{eqnarray}
&&
i \biggl( \frac{\partial A}{\partial t} + C_x
\frac{\partial A}{\partial x} + C_y
\frac{\partial A}{\partial y} \biggr) + \alpha \frac{\partial^2
A}{\partial x^2} + \beta \frac{\partial^2
A}{\partial y^2} + \gamma \frac{\partial^2
A}{\partial x \partial y}
\nonumber \\ &&\qquad
- \xi \,|A^2|  A - 2 \zeta \,|B|^2  A = 0 \, ,
\end{eqnarray}
and
\begin{eqnarray}
&&
i \biggl( \frac{\partial B}{\partial t} + C_x
\frac{\partial B}{\partial x} - C_y
\frac{\partial B}{\partial y} \biggr) + \alpha \frac{\partial^2
B}{\partial x^2} + \beta \frac{\partial^2
B}{\partial y^2} - \gamma \frac{\partial^2
B}{\partial x \partial y}
\nonumber \\ && \qquad
- \xi \,|B^2|  B - 2 \zeta \,|A|^2 B  = 0 \, ,
\end{eqnarray}
\label{CNLSE}
\end{subequations}
where $A$ and $B$ are the amplitudes of the slowly varying 
wave envelopes such that the surface elevations (in meters) are 
given by $\eta_A=(1/2)A({\bf r},t)\exp(ikx+ily-i\omega t)$+c.c.~and
$\eta_B=(1/2)B({\bf r},t)\exp(ikx-ily-i\omega t)$+c.c.~where c.c.~denotes complex
conjugate. Here, $C_x=\omega k/2\kappa^2$ and $C_y=\omega l/2\kappa^2$ are the group
speed components along the $x$ and  $y$ axis, respectively, 
and the group velocity dispersion coefficients are $\alpha=\omega(2 l^2-k^2)/8\kappa^4$,
$\beta=\omega (2k^2-l^2)/8\kappa^4$ and $\gamma=-3\omega l k /4\kappa^4$, 
and the nonlinearity coefficients are given by $\xi=\omega \kappa^2/2$ and 
$\zeta=\omega(k^5-k^3l^2-3kl^4-2k^4\kappa+2k^2 l^2\kappa+2l^4\kappa)/2\kappa^2(k-2\kappa)$ 
(see Ref. \cite{r3}). It should be pointed out that Eq.\ (1) was first written down about 
forty years ago by Benney and Newell \cite{bn} who gave formulae for all the coefficients 
except $\xi$ and $\zeta$, which however can be found in Ref.\ \cite{r3}.

It is easy to verify that the system of two-dimensional CNLS equations
(\ref{CNLSE}) has the space independent harmonic solutions $A_{eq} = A_0
\, \exp[- i (\xi |A_0|^2 + 2 \zeta |B_0|^2 ) t]$ and $B_{eq} = B_0 \,
\exp[- i (\xi |B_0|^2 + 2 \zeta |A_0|^2 ) t]$. 
Assuming a small (linear) harmonic perturbation around the 
above mentioned equilibrium states with the wavevector 
$\mathbf{K} = (K, L)$ and the frequency $\Omega$, i.e.\ substituting with 
$A \rightarrow (A_0 + \epsilon A_1) \exp [-i(\xi |A_0|^2 + 2 \zeta |B_0|^2 ) t]$ and 
$B \rightarrow (B_0 + \epsilon B_1) \exp [- i (\xi |B_0|^2 + 2 \zeta |A_0|^2 ) t]$
into (1), linearizing in the small real parameter $\epsilon \ll 1$, 
then separating the real and imaginary parts, combining the resultant 
equations, and Fourier transforming, we obtain the nonlinear dispersion relation 
\begin{equation}
\bigl[ (\Omega - C_x K - C_y L)^2 - \Omega_1^2 \bigr] \,
\bigl[ (\Omega - C_x K + C_y L)^2 - \Omega_2^2 \bigr] = \Omega_{c}^4 \, ,
\label{DR}
\end{equation}
where
$ \Omega_1^2 \, = (\alpha K^2 + \beta L^2 - \gamma K L)
(\alpha K^2 + \beta L^2 + \gamma K L + 2 \xi  |A_0|^2)$,
$\Omega_2^2 \, = (\alpha K^2 + \beta L^2 + \gamma K L)
(\alpha K^2 + \beta L^2 - \gamma K L + 2 \xi |B_0|^2)$,
and
$\Omega_c^4 \, = 16 \, \zeta^2 \, |A_0|^2 |B_0|^2 \,
(\alpha K^2 + \beta L^2 - \gamma K L) (\alpha K^2 + \beta L^2 + \gamma K L)$.
For one-dimensional wave propagation, i.e. for  $L=0$, Eq.\ (\ref{DR}) is identical to 
Eq.\ (11) of Ref.\ \cite{r3}.  Onorato {\it et al.}\ \cite{r3} presented numerical results 
for the modulational instability regimes and growth rates for $L = 0$ and $A_0 = B_0$ from 
their Eq.\ (11). We note that the dispersion relation (2) is nonlinear in the wave amplitudes 
(but linear in the small expansion parameter).

Taking into account that a measure of the wave nonlinearities are given by the
``steepnesses'' $\kappa A$ and $\kappa B$, we will use the natural 
normalizations $A_0= A_0'/\kappa$ and $B_0= B_0'/\kappa$ for the wave amplitudes.
For the wavenumbers and frequencies we use $K'=K/\kappa$, $L'=L/\kappa$, 
$k'=k/\kappa$, $l'=l/\kappa$, and $\Omega'=\Omega/\omega$.  The coefficients are 
normalized as $C_x'=C_x\kappa/\omega=k'/2$, $C_y'=C_y\kappa/\omega=l'/2$,
$\alpha'= \alpha\kappa^2 /\omega=(2l'^2-k'^2)/8$, 
$\beta'=\beta\kappa^2/\omega=(2k'^2-l'^2)/8$, 
$\gamma'=\gamma \kappa^2/\omega=-3l'k'/4$,
$\xi'=\xi/\omega k^2=1/2$, and
$\zeta'=\zeta/\omega k^2=[(k')^5-(k')^3 (l')^2-3k'(l')^4-2 (k')^4+2(k')^2 (l')^2+2(l')^4]/2(k'-2)$,
and all quantities in Eqs. (2) and (3) can be replaced with their primed counterparts.
Here we have $k'=\cos \theta$ and $l'=\sin \theta$, where $\theta$ is the
angle between the wave directions and the dichotome.
Except for numerical factors, Eqs. (2) and (3) then only contain the angle
$\theta$, the known wavenumbers $K'$ and $L'$, the wave amplitudes $A_0'$ and $B_0'$, 
and the unknown frequency $\Omega'$.

In the following, we numerically solve  our nonlinear dispersion relation (\ref{DR}) 
and present the growth rate $\Gamma$ (the imaginary part of $\Omega$) in Figs.\ 1--3, 
where we have studied the impact of different angles $\theta$ on the growth rates 
for interacting waves. For all cases we used the waves amplitudes
$0.1/\kappa$ for the two waves. In the left-hand panels of Figs. 1 and 2
we show the single wave cases, which exhibit the standard
Benjamin-Feir instability [cf. Eq. (3.6) and Fig. 1 of Ref. \cite{Segur05}],
tilted by the angle $\theta$ in the $(K,L)$ plane. The right-hand panels
show the cases of interacting waves. We see from Fig. 1 that a
relatively small $\theta=\pi/8$ gives rise to a new instability with a 
maximum growth rate that is more than twice as large as the ones for the single
wave cases, in the direction of the dichotome. For a larger angle $\theta=\pi/4\approx 0.79$, displayed
in Fig. 2, we see that the two waves do not interact to enhance the linear growth rate significantly.
We note (as pointed out in Ref. \cite{r3}) that the coefficient $\alpha$ changes sign when
$\theta=\arctan(l/k)=\arctan(1/\sqrt{2})\approx 0.615$ rad $\approx35.3^\circ$ so that we have 
a focusing (defocusing) instability along the $x$ axis for $\theta < 0.615$ ($>0.615$). This clearly
stabilizes the waves so that they only exhibit the standard Benjamin-Feir
instability but do not interact to enhance the instability.
On the other hand, for two counter-propagating waves with $\theta=\pi/2$, we see in Fig. 3
that there is again an instability along the dichotome (in agreement with Ref. \cite{r3}) 
and also obliquely in narrow bands almost perpendicular to the two waves.

In order to study the dynamics of nonlinearly interacting water wave 
packets, we solve the coupled system of equations (\ref{CNLSE}). 
The results are displayed in Fig.\ 4.
In the numerical simulation, we have used the normalization
$A' = A/\kappa$, $B'= B/\kappa$, $t' = \omega t$, $x' = \kappa x$, 
and $y'= \kappa y$ (the other scaled parameters
are as above), while the results are shown in dimensional units in Fig. 4. 
We have used the same parameters as in the right-hand panel of Fig. 1, where
the two interacting waves initially have the amplitude $A=B=0.1/\kappa$,
plus low-amplitude noise (random numbers) of the order $10^{-3}/\kappa$ to 
give a seed for any instability.
The wavefronts in Fig.\ 4 are initially directed primarily in the $x$
direction, reflecting the maximum growth rate in the $K$ direction
in the right-hand panel of Fig.\ 1.  The linear growth saturates
in the formation of wavepackets, localized in the $x$ direction,
that are strongly correlated
between wave $A$ and $B$; see the panels at time $t=900/\omega$. At later
times the waves $A$ and $B$ decouple and there are scattered large-amplitude
waves that are localized both in the $x$ and $y$ directions. In order to compare
with the case of a single wave we have also run a simulation where $B$ was set to zero, so
that we have the standard Benjamin-Feir instability shown in the upper left
panel of Fig 1. In this case (not shown here), we could not see the formation
of well-defined wave-packets, but the instability gave rise to a wide
spectrum of waves in different directions, in agreement with the linear 
analysis in the left-hand panels of Fig. 1. The new instability due to the coupling of the two
waves, shown in the right-hand panel of Fig. 1, has a well-defined maximum
in the $x$ direction, which is important for the localization of 
wave energy into localized wavepackets seen in Fig. 4. Taking some
typical data from ocean waves \cite{Hasselmann76} where a typical
wave frequency is 0.09 Hz, we have $\omega=0.56\,{\rm s}^{-1}$, and
$\kappa=\omega^2/g\approx0.033\,{\rm m}^{-1}$. The waves $A$ and $B$ in
Fig. 4, then have the initial amplitudes $|A|=|B|=0.1/\kappa\approx 3$ meters.
In Fig. 4, we see at $t=1200/\omega$ ($\approx 670$ seconds) that wave
$A$ has some localized wave packets with 
a maximum amplitude of $\approx 0.35/\kappa\approx 10$ meters. 

\begin{figure}
\includegraphics[width=0.45\textwidth]{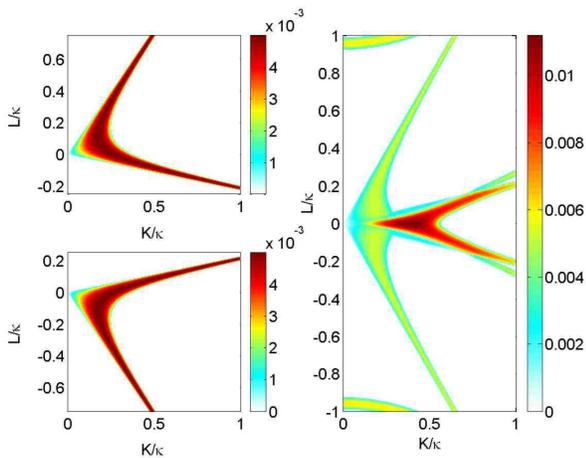}
\caption{ The normalized growth rate $\Gamma/\omega$ plotted as a function
of $K/\kappa$ and $L/\kappa$. Here we have used $\theta=\pi/8$
and the wave amplitudes $A_0=B_0=0.1/\kappa$. The left upper and lower
panels show the cases with a single wave $A_0$ and $B_0$, respectively,
while the right panel shows the case of interacting waves.
}
\end{figure}

\begin{figure}
\includegraphics[width=0.45\textwidth]{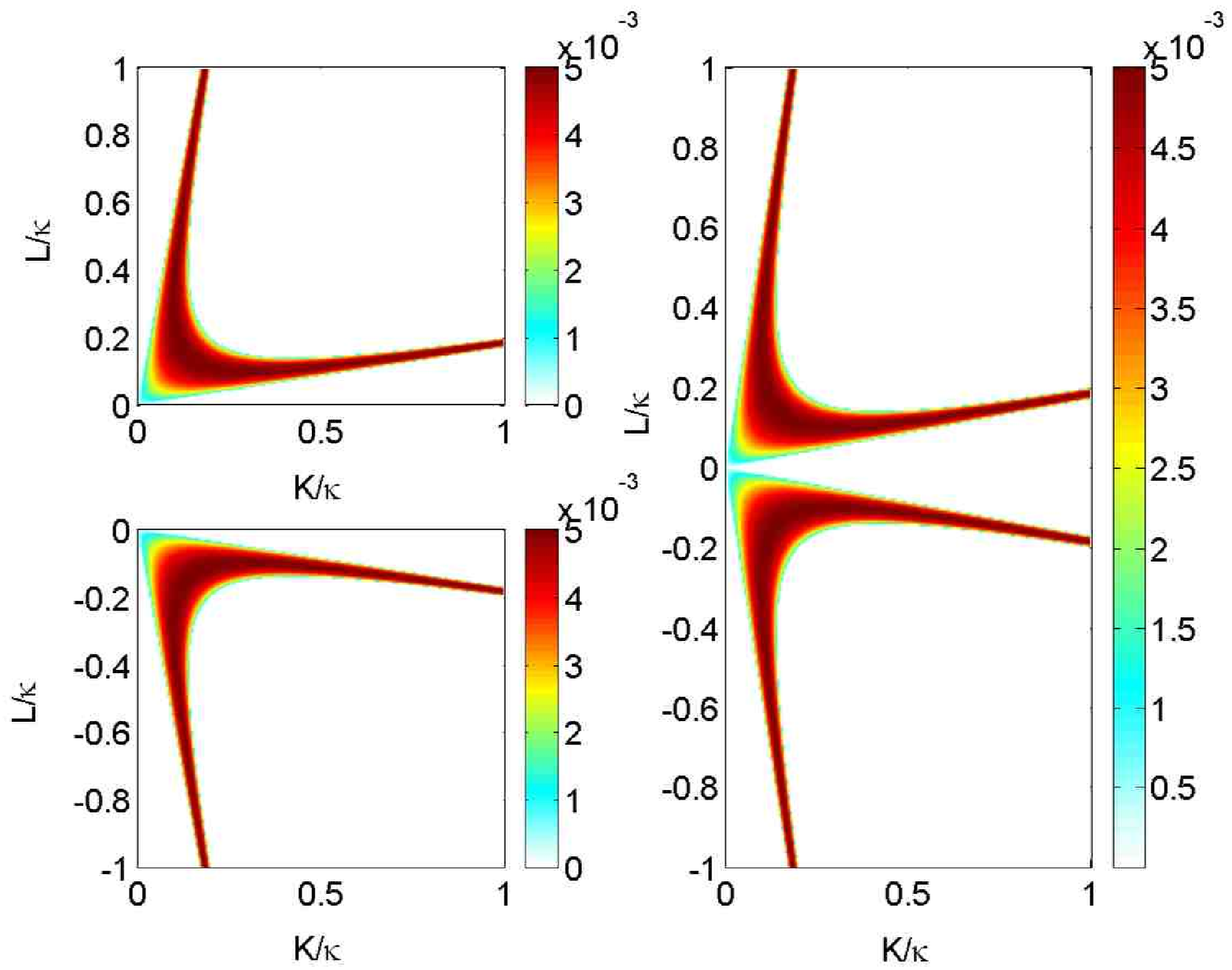}
\caption{ The normalized growth rate $\Gamma/\omega$ plotted as a function
of $K/\kappa$ and $L/\kappa$. Here we have used $\theta=\pi/4$
and the wave amplitudes $A_0=B_0=0.1/\kappa$. The left upper and lower
panels show the cases with a single wave $A_0$ and $B_0$, respectively,
while the right panel shows the case of interacting waves.
}
\end{figure}

\begin{figure}
\includegraphics[width=0.45\textwidth]{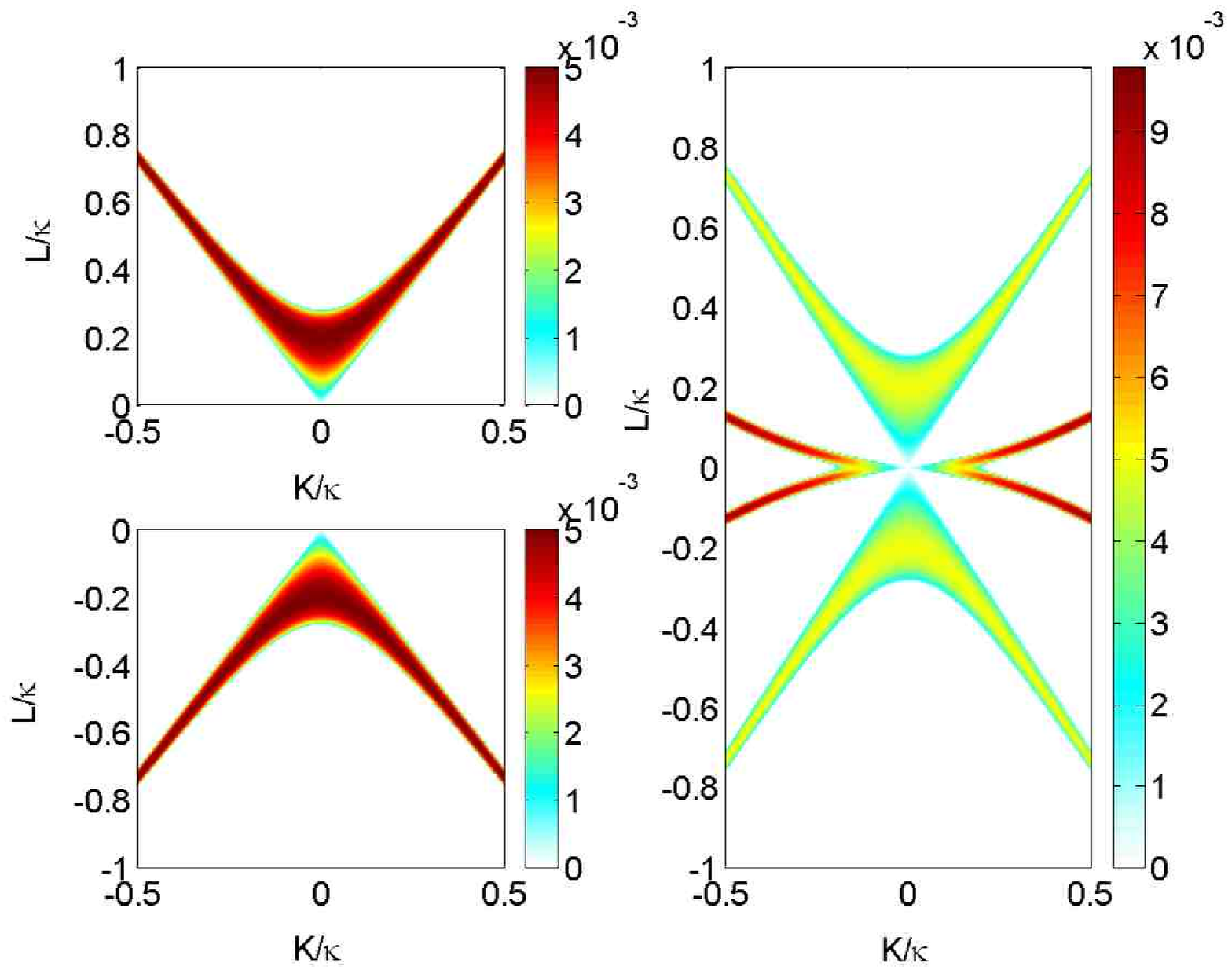}
\caption{ The normalized growth rate $\Gamma/\omega$ plotted as a function
of $K/\kappa$ and $L/\kappa$. Here we have used $\theta=\pi/2$
and the wave amplitudes $A_0=B_0=0.1/\kappa$. The left upper and lower
panels show the cases with a single wave $A_0$ and $B_0$, respectively,
while the right panel shows the case of interacting waves.
}
\end{figure}

\begin{figure*}
\includegraphics[width=0.8\textwidth]{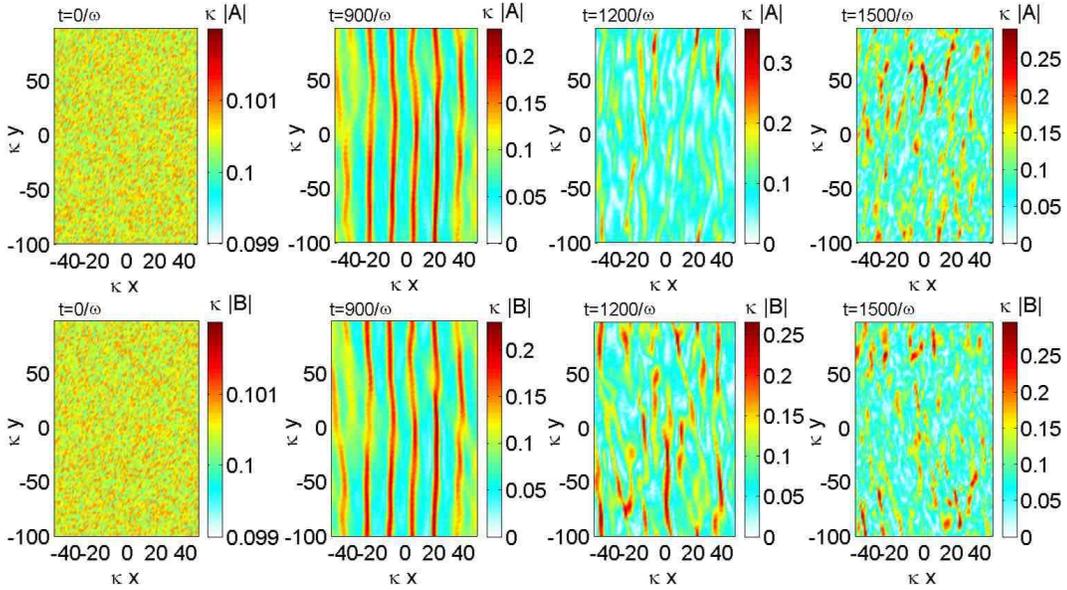}
\caption{ The interaction between two waves, initially with equal amplitudes
$|A|=|B|=0.1\,\kappa^{-1}$ and a propagation angle of $\theta=\pi/8$ relative
to the dichotome. Added to the initially homogeneous wave envelopes is a low-amplitude
noise of order $10^{-3}/\kappa$ to give a seed to the modulational instability.
}
\end{figure*}

To summarize, we have presented a theoretical study of the modulational
instabilities of a pair of nonlinearly interacting two-dimensional waves 
in deep water, and have shown that the full dynamics of these interacting waves
gives rise to localized large-amplitude wavepackets.
Starting from the CNLS equations of Onorato {\it et al.}\ \cite{r3}, we
have derived a nonlinear dispersion equation. The latter has been numerically
analyzed to show the dependence of obliqueness of the interacting waves and of the
modulations on new classes of modulational 
instabilities that we found in multi-dimensional situations. Furthermore, the numerical 
analysis of the full dynamical system reveals that even waves that are separately 
modulationally stable can, when nonlinear interactions are taken into account, 
give rise to novel behavior such as the formation of large-amplitude coherent wave packets with
amplitudes more than three times the ones of the initial waves. This behavior is
very different from that of a single wave which experiences the standard Benjamin-Feir
instability which dissolves into a wide spectrum of waves.
These results will have relevance to the nonlinear instability of colliding water waves, 
which may interact nonlinearly in a constructive way to produce large-amplitude 
freak waves in the oceans.

\acknowledgments

This research was partially supported by the Swedish Research Council and
the Deutsche Forschungsgemeinschaft.


\end{document}